\documentclass[11pt,letterpaper]{article}
\usepackage{graphicx}

\newcommand{\re}{{\bf r}  }
\newcommand{\rp}{{\bf r'}  }
\newcommand{\be}{\begin{equation}  }
\newcommand{\ee}{\end{equation}  }

\newcommand{\nup}{{n_\uparrow}  }
\newcommand{\ndo}{{n_\downarrow}  }

\begin{document}

\title{Two Avenues to Self-Interaction Correction within Kohn-Sham
Theory: Unitary Invariance is the Shortcut} 

\author{Stephan K\"ummel and John P.\ Perdew \\
Department of Physics and Quantum Theory Group\\ 
Tulane  University\\
New Orleans, Louisiana 70118, USA}

\maketitle

\begin{abstract}

The most widely-used density functionals for the exchange-correlation
energy are inexact for one-electron systems. Their self-interaction
errors can be severe in some applications. The problem is not only to
correct the self-interaction error, but to do so in a way that will
not violate size-consistency and will not go outside the standard
Kohn-Sham density functional theory. The solution via the optimized
effective potential  
(OEP) method will be discussed, first for the Perdew-Zunger
self-interaction correction (whose performance for molecules is
briefly summarized) and then for the more modern
self-interaction corrections based upon unitarily-invariant indicators
of iso-orbital regions. For the latter approaches, the OEP
construction is greatly simplified. The kinetic-energy-based iso-orbital
indicator $\tau^W_\sigma(\re)/\tau_\sigma(\re)$ will be discussed and
plotted, along with an alternative exchange-based indicator. 

\end{abstract}

\section{Introduction}

Paradoxically, some of the most widely-used and reliable theoretical
approaches to many-electron systems, including the local spin density
\cite{lsd} (LSD) and generalized gradient \cite{gga} (GGA) approximations, are
not exact for one-electron systems. The error they make in these
systems is called the self-interaction error. An early
self-interaction correction \cite{sic,sic2} (SIC) to LSD led to
dramatic successes and failures, and has been 
largely bypassed by the development of GGA, which provides a more uniform
improvement over LSD and has an easier Kohn-Sham theory implementation. Since
the time finally seems ripe for the development of reliable
self-interaction-free approximations, this article will discuss some of the
possibilities for self-interaction correction within Kohn-Sham density
functional theory \cite{lsd}.

In Kohn-Sham theory, the many-electron ground-state spin densities
$\nup(\re)$ and $\ndo(\re)$ and energy $E$ are predicted by self-consistent
solution of the one-electron equations
\be
\label{kseq}
\left[-\frac{\hbar^2}{2m}\nabla^2 + v_s^\sigma(\re)\right]
\varphi_{\alpha\sigma}(\re) 
=\varepsilon_{\alpha\sigma}\varphi_{\alpha\sigma}(\re),
\ee
\be
\label{defn}
n_\sigma(\re)=\sum_\alpha|\varphi_{\alpha\sigma}|^2
\Theta(\mu-\varepsilon_{\alpha\sigma}) 
\ee
where the spin-dependent effective potential is
\be
\label{vks}
v_s^\sigma(\re)=v_\sigma(\re)+e^2 \int d^3r' \, \frac{n(\rp)}{|\re-\rp|}
+ v_\mathrm{xc}^\sigma(\re).
\ee
In Eq.\ (\ref{vks}), $v_\sigma(\re)$ is the external potential created by the
nuclei and external scalar fields, $n=\nup+\ndo$ is the total electron
density, and
\be
v_\mathrm{xc}^\sigma(\re)=\frac{\delta E_\mathrm{xc}[\nup,\ndo]}{\delta
  n_\sigma(\re)} 
\ee
is the exchange-correlation potential. $\mu$ is the Fermi level, and
$\Theta(x)$ is the step function ($\Theta=1$ for $x>0$ and $\Theta=0$ for
$x < 0$.) The energy is
\be
E=T_s[\nup,\ndo]+\sum_\sigma \int d^3r \, n_\sigma(\re) v_\sigma(\re)
+ \frac{e^2}{2} \int d^3r \, \int d^3r' \frac{n(\re)n(\rp)}{|\re-\rp|}
+E_\mathrm{xc}[\nup,\ndo].
\ee
The non-interacting kinetic energy is
\be
T_s[\nup,\ndo]=\sum_\sigma \int d^3r \, t_s(\re)=
\sum_\sigma \int d^3r \, \tau_\sigma
(\re),
\ee
where
\be
\label{ts}
t_\sigma(\re)=\sum_\alpha \varphi_{\alpha \sigma}^*(\re)
\left(-\frac{\hbar^2}{2m} \nabla^2 \right) \varphi_{\alpha \sigma}(\re)
\Theta(\mu-\varepsilon_{\alpha\sigma}),
\ee
\begin{eqnarray}
\tau_\sigma(\re)&=&
\sum_\alpha \frac{\hbar^2}{2m} 
|\nabla \varphi_{\alpha \sigma}(\re) |^2
\Theta(\mu-\varepsilon_{\alpha\sigma})
\nonumber \\
&=& t_\sigma(\re)+\frac{\hbar^2}{4m}\nabla^2 n_\sigma(\re).
\label{tau}
\end{eqnarray}
Since the Kohn-Sham orbitals 
$\varphi_{\alpha \sigma}(\re)$ are functionals \cite{lsd} of the spin
densities $\nup$ and $\ndo$, so is $T_s$.

While the equations of the previous paragraph are exact in principle, in
practice
$E_\mathrm{xc}[\nup,\ndo]=E_\mathrm{x}[\nup,\ndo]+E_\mathrm{c}[\nup,\ndo]$
must be 
approximated. In the local spin density approximation \cite{lsd}, 
\be
E_\mathrm{xc}^\mathrm{LSD}[\nup,\ndo]=
\int d^3r \, n \epsilon_\mathrm{xc}^\mathrm{unif}(\nup,\ndo), 
\ee
where $\epsilon_\mathrm{xc}^\mathrm{unif} $ is the known exchange-correlation
energy of an electron gas with uniform spin densities $\nup$, $\ndo$.
In the generalized gradient approximation \cite{gga},
\be
E_\mathrm{xc}^\mathrm{GGA}[\nup,\ndo]=\int d^3r \, n
\epsilon_\mathrm{xc}^\mathrm{GGA}(\nup,\ndo,\nabla \nup,\nabla \ndo).
\ee
These approximations are exact for a uniform density and accurate for a
slowly-varying $n(\re)$, but are not exact for one electron densities.

One-electron densities are fully spin-polarized (e.g., $\nup=n$ and
$\ndo=0$). In a one electron system, the exchange energy must cancel the
Hartree electrostatic energy:
\be
\label{onex}
E_\mathrm{x}[\nup,0]=-\frac{e^2}{2} \int d^3r \,
\frac{\nup(\re)\nup(\rp)}{|\re-\rp|} 
\hspace{5mm}(N=N_\uparrow=1)
\ee
and the correlation energy must vanish:
\be
\label{onec}
E_\mathrm{c}[\nup,0]= 0
\hspace{5mm}(N=N_\uparrow=1).
\ee
To satisfy Eqs.\ (\ref{onex}) and (\ref{onec}), Perdew and Zunger \cite{sic}
proposed a self-interaction correction to LSD:
\be
\label{sicfunc}
E_\mathrm{xc}^\mathrm{SIC}=E_\mathrm{xc}^\mathrm{LSD}[\nup,\ndo]-
\sum_{\alpha \sigma}\left\{
\frac{e^2}{2} \int d^3r \int d^3r' \,
\frac{n_{\alpha \sigma}(\re)n_{\alpha \sigma}(\rp)}{|\rp-\re|}
+ E_\mathrm{xc}^\mathrm{LSD}[n_{\alpha \sigma},0]\right\},
\ee
\be
\label{sicpot}
v_\mathrm{xc}^{\mathrm{SIC},\alpha \sigma}(\re)=
v_\mathrm{xc}^{\mathrm{LSD},\sigma}([\nup,\ndo];\re)-
e^2 \int d^3r' \, \frac{n_{\alpha \sigma}(\rp)}{|\rp-\re|}
-v_\mathrm{xc}^{\mathrm{LSD},\uparrow}([n_{\alpha\sigma},0];\re),
\ee
where
\be
\label{norb}
n_{\alpha\sigma}(\re)=|\varphi_{i\sigma}(\re)|^2
\Theta(\mu-\varepsilon_{\alpha\sigma}) 
\ee
is an orbital density. The potential (\ref{sicpot}) has the correct asymptotic
behavior
\be
v_\mathrm{xc}^\sigma(\re)\rightarrow -\frac{e^2}{r}
\hspace{5mm} \mbox{as} \hspace{5mm} r \rightarrow \infty
\ee
as one moves away from any compact system, while
$v_\mathrm{xc}^{\mathrm{LSD},\sigma}(\re)$ tends to zero exponentially in this
limit. Unlike LSD, the SIC exchange-correlation energy displays
\cite{sic2} a derivative
discontinuity very much like that of the exact $E_\mathrm{xc}[\nup,\ndo]$.

There is no unique way to make a self-interaction correction, and alternatives
to Eqs.\ (\ref{sicfunc}) -- (\ref{norb}) have been proposed
\cite{cortona,nesbet,whitehead,lundin,unger}. 
But Eqs.\ (\ref{sicfunc}) -- 
(\ref{norb}) have been widely tested for atoms
\cite{sic,gunnarsson,harrison,krieger,chen}, atomic ions \cite{cole}, molecules
\cite{pederson1,pederson2,johnson,goedecker,jp,garza,patch1,patch2,polo1,polo2,polo3}
and solids 
\cite{strange,tem} (see earlier references in Ref.\ \cite{jp}). SIC is 
exact for one-electron systems, and usually accurate for strongly localized
electrons. For covalent molecules near equilibrium, it has been argued
\cite{polo1,polo2,polo3} that the self-interaction
{\it error} 
in LSD and GGA exchange \cite{handy} is needed to mimic the effect of
static correlation on the electron density.

In applications to molecules, the performance of SIC is somewhat
mixed. Total energies are better than in LSD, and the highest occupied
orbital energy is much closer to minus the ionization potential than
in LSD \cite{pederson2,goedecker}. 
The localized SIC valence orbitals correspond to the localized bonds
and lone pairs of chemical intuition \cite{goedecker}.
SIC significantly improves the
energy barriers to chemical reactions \cite{johnson,patch2}, but net
reaction energies are less strongly improved relative to LSD
\cite{patch2}.
Many nuclear magnetic resonance properties of molecules are improved
by SIC \cite{patch1}. There are relatively few studies of atomization
energies in SIC, but there seems to be an improvement over LSD for the
cases studied: $\mathrm{Li}_2$ \cite{pederson2},  $\mathrm{O}_2$
\cite{patch1}, and $\mathrm{N}_2$ \cite{polo2}. The most
disappointing results are the SIC bondlengths, which are shorter than
the experimental ones by $\approx$ 0.07 bohr on average
\cite{goedecker}, while the LSD bond lengths are much more realistic.

\section{Perdew-Zunger SIC within Kohn-Sham Theory}

The SIC of Eqs.\ (\ref{sicfunc}) -- (\ref{norb}) goes outside the
Kohn-Sham scheme by introducing an {\it orbital-dependent} effective
potential $v_s^{\alpha\sigma}(\re)$. As a result, the self-consistent
SIC orbitals are not Kohn-Sham orbitals, and are not even strictly
orthogonal unless off-diagonal Lagrange multipliers are
introduced. The SIC orbitals tend to localize around atomic centers,
while the Kohn-Sham orbitals are delocalized canonical or molecular
orbitals. 
The SIC orbitals can be found, even for molecules, by directly
minimizing Eq.\ (\ref{sicfunc}) under the constraint of orbital
orthogonality \cite{pederson1,pederson2,polo1,polo2,polo3}. 

Although
not a Kohn-Sham theory, the Perdew-Zunger SIC belongs \cite{sic} to a
wider class of density functional theories. At least to the extent
that the SIC orbitals are localized, it is also a size-consistent
theory \cite{sic2}, i.e., one which works consistently well for small
or large systems.   

But there are clearly computational and conceptual advantages to
Kohn-Sham theory, not only for the ground state but also for
time-dependent processes and excitations. To bring SIC under the
umbrella of Kohn-Sham theory, one must construct a common effective
potential for all the occupied orbitals of spin $\sigma$. Especially
in the context of time-dependent DFT, different procedures emphasizing
computational simplicity have been suggested to construct a common
local potential \cite{ullrich,chu1,madjet}, and the influence of the
self-interaction correction on optical properties of atoms
\cite{chu1}, the molecule $\mathrm{N}_2$ \cite{chu2} and clusters
\cite{ullrich,app} has been discussed.   
A rigorous way of constructing a common potential is given by the
optimized effective potential (OEP) 
method \cite{talman,yang,oep}. For any orbital functional
$E[\{\varphi_{\alpha\sigma}\}]$, the OEP method delivers a Kohn-Sham
potential and a set of Kohn-Sham orbitals which minimize that
functional. When the orbital functional is Hartree-Fock, there is no
problem, but when it is SIC (Eq.\ (\ref{sicfunc})) the resulting scheme
is not size-consistent: Applied to one atom, where all the Kohn-Sham
orbitals are localized, this scheme will deliver a properly
self-interaction-corrected energy. But, applied to a periodic lattice
of atoms separated by large lattice constants, where all the Kohn-Sham
orbitals are delocalized, this scheme will produce no self-interaction
correction to the energy of an atom, since the sum in Eq.\
(\ref{sicfunc}) will then vanish on a per-atom basis \cite{sic}.
The considerations put forward in Refs.\ \cite{sic,sic2} suggest that this
is true for all Perdew-Zunger-like SIC schemes that directly use the Kohn-Sham
orbitals. Therefore, such   
schemes would be good for atoms, but would degrade for molecules or
clusters as the number of atoms increased.

A clue to the solution of this problem was given in the work of
Pederson, Heaton and Lin \cite{pederson2}, who introduced {\it two}
sets of occupied orthonormal orbitals related by unitary
transformation: the localized SIC orbitals, and the delocalized
canonical orbitals. Garza, Nichols and Dixon \cite{garza} proposed
that the canonical orbitals could be Kohn-Sham orbitals belonging to
an optimized effective potential $v_s^\sigma(\re)$ constructed from
the localized orbitals. In their work, and in that of Patchovski and
Ziegler \cite{patch1,patch2}, the Krieger-Li-Iafrate approximation
\cite{kli} to OEP is used, as is a standard (non-optimal) localizing
transformation. 

As an exactification of this approach, the correct Kohn-Sham version
of Perdew-Zunger SIC would be conceptually this: Start with a given
external potential $v_\sigma(\re)$ and electron number $N$. Form a
trial effective potential $v_s^\sigma(\re)$, and solve Eq.\
(\ref{kseq}) to find the corresponding occupied Kohn-Sham
orbitals. Then find the unitary transformation to localized orbitals
that minimizes Eq.\ (\ref{sicfunc}). Finally, choose the effective
potential that delivers the lowest minimum of Eq.\ (\ref{sicfunc}). 

\section{Unitarily Invariant Iso-Orbital Indicators}

The prescription outlined above for the implementation of the
Perdew-Zunger self-interaction correction to LSD (or GGA) within
Kohn-Sham theory was greatly complicated by the fact that the
self-interaction correction was not invariant under a unitary
transformation of the occupied orbitals. This section will discuss
self-interaction corrections that {\it are} unitarily  invariant, and
thus can be implemented within Kohn-Sham theory by a direct
application of the OEP method \cite{talman,yang,oep} to the Kohn-Sham
orbitals. This subject is timely because of the recent appearance of
accurate and efficient solutions \cite{yang,oep} to the OEP problem.

A Slater determinant of occupied orbitals of a given spin $\sigma$ is
invariant under unitary transformation of those orbitals, and so is
any quantity that can be constructed from the Slater determinant, such
as the spin density of Eq.\ (\ref{defn}) or the kinetic energy
densities of Eqs.\ (\ref{ts}) and (\ref{tau}). The one-electron
density matrix
\be
\rho_\sigma(\re,\rp)=\sum_\alpha 
\varphi_{\alpha\sigma}^*(\re) \varphi_{\alpha\sigma}(\rp)
\Theta(\mu-\varepsilon_{\alpha\sigma})
\ee
is also invariant. (The step function must of course be re-interpreted
as a restriction to the occupied orbital space.) The exact exchange
energy
\be
\label{exx}
E_\mathrm{x}=-\frac{e^2}{2}\sum_\sigma \int d^3r \,\int d^3 r' \, 
\frac{\rho_\sigma^2(\re,\rp)}{|\rp-\re|} 
\ee
is clearly invariant, as is the the local exchange energy per electron
$e_\mathrm{x}(\re)$:
\be
\label{ex}
e_\mathrm{x}(\re)=-\frac{e^2}{2}\sum_\sigma \int d^3r' \, 
\frac{\rho_\sigma^2(\re,\rp)}{n(\re)|\rp-\re|}.
\ee

On the ``Jacob's Ladder'' \cite{jaclad} of density functional
approximations, full 
freedom from self-interaction error is achieved only at the hyper-GGA
level, which employs full exact exchange and a highly nonlocal
functional of the occupied orbitals for correlation. A somewhat
different way to eliminate the self-interaction error is via a local
hybrid functional \cite{jaram}. But in either case one needs an {\it
iso-orbital indicator} to identify regions of space in which the
electron density is dominated by a single orbital shape. The
iso-orbital regions where $\nup \ndo=0$ 
are one-electron regions in which the correlation energy per electron
$e_\mathrm{c}(\re)$ can and should be zeroed out by a
self-correlation-free density functional.

The exact exchange energy of Eq.\ (\ref{exx}) is self-interaction free,
since for a one-electron ($N=N_\sigma=1$) ground-state 
$\rho_\sigma(\re,\rp)=n_\sigma^{1/2}(\re) \, n_\sigma^{1/2}(\rp)$. Thus
\be
y_\sigma(\re,\rp)=
\frac{n_\sigma^{1/2}(\re)n_\sigma^{1/2}(\rp)}{\rho_\sigma(\re,\rp)}
\ee
is an iso-orbital indicator which equals unity when both $\re$ and
$\rp$ are in an iso-orbital region. However, as $\rp \rightarrow \re$, 
$y_\sigma(\re,\rp)$ tends to 1 in any region, iso-orbital or
not. This problem does not arise for
\be
\label{defx}
x_\sigma(\re)=\lim_{\rp\rightarrow \re}
\frac{\nabla_\re \cdot \nabla_\rp 
n_\sigma^{1/2}(\re)n_\sigma^{1/2}(\rp)}
{\nabla_\re \cdot \nabla_\rp \rho_\sigma(\re,\rp)}
=\frac{\tau_\sigma^W(\re)}{\tau_\sigma(\re)}.
\ee
Eq.\ (\ref{defx}) provides a point-by-point iso-orbital indicator
which equals unity in 
any iso-orbital region and is otherwise bounded between 0 and 1
\cite{kurth}. In Eq.\ (\ref{defx}), $\tau_\sigma(\re)$ is the kinetic
energy density of Eq.\ (\ref{tau}), and
\be
\label{tauw}
\tau_\sigma^W(\re)=\frac{\hbar^2}{8m}\frac{|\nabla
n_\sigma(\re)|^2}{n_\sigma(\re)} 
\ee
is the von Weizs\"acker or bosonic kinetic energy density. For a uniform
density, $x_\sigma(\re)$ vanishes everywhere. 

$x_\sigma(\re)$ of Eq.\ (\ref{defx}) is clearly invariant under
unitary transformation of the occupied orbitals. The idea of using the
condition $\tau_\sigma(\re)=\tau_\sigma^W(\re)$ to identify an
iso-orbital region and zero out the self-correlation goes back to
Colle and Salvetti \cite{colle}, but in density functional theory to
Becke \cite{becke} and Dobson \cite{dobson}. $x_\sigma(\re)$ is an
ingredient of self-correlation free meta-GGA's including those of
Refs.\ \cite{pkzb} and \cite{newmgga}, and of local hybrids
\cite{jaram} and hyper-GGA's \cite{jaclad}.

\begin{figure}[h]
\includegraphics[width=8.1cm]{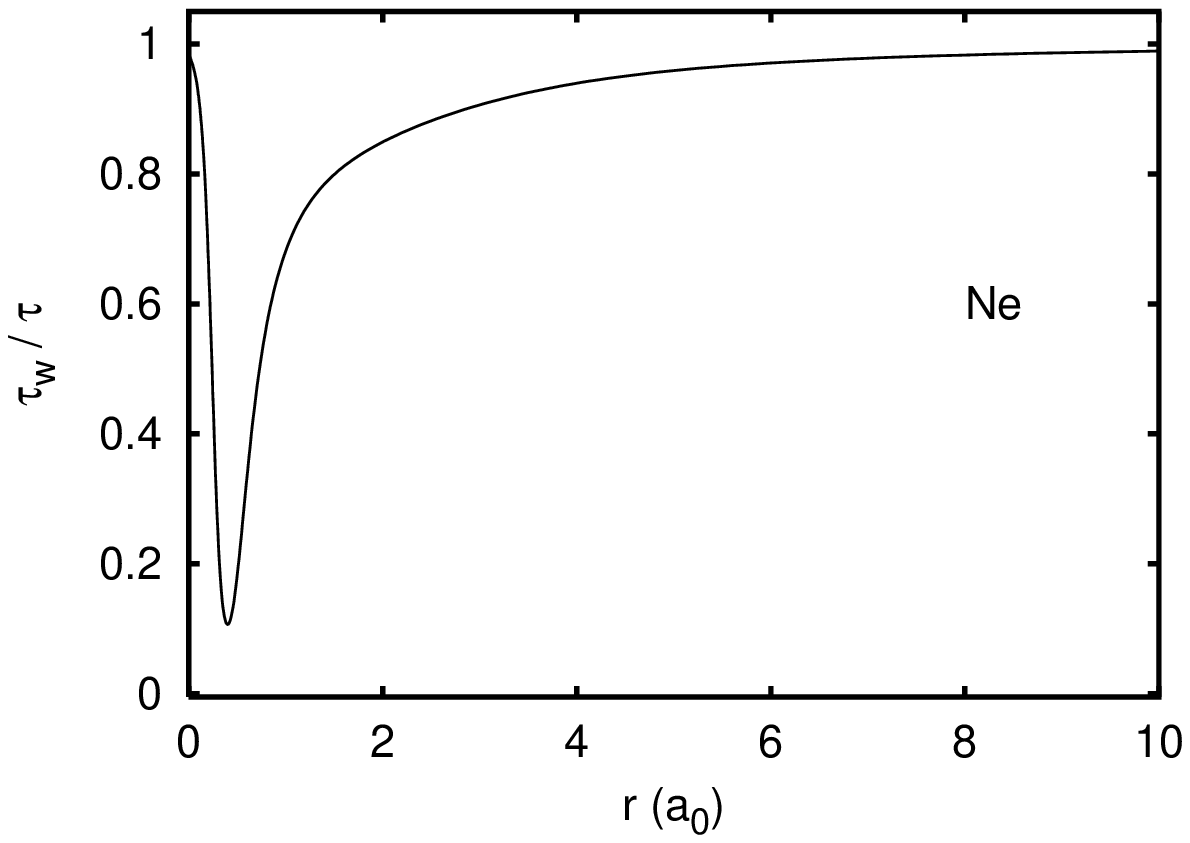}
\includegraphics[width=8.1cm]{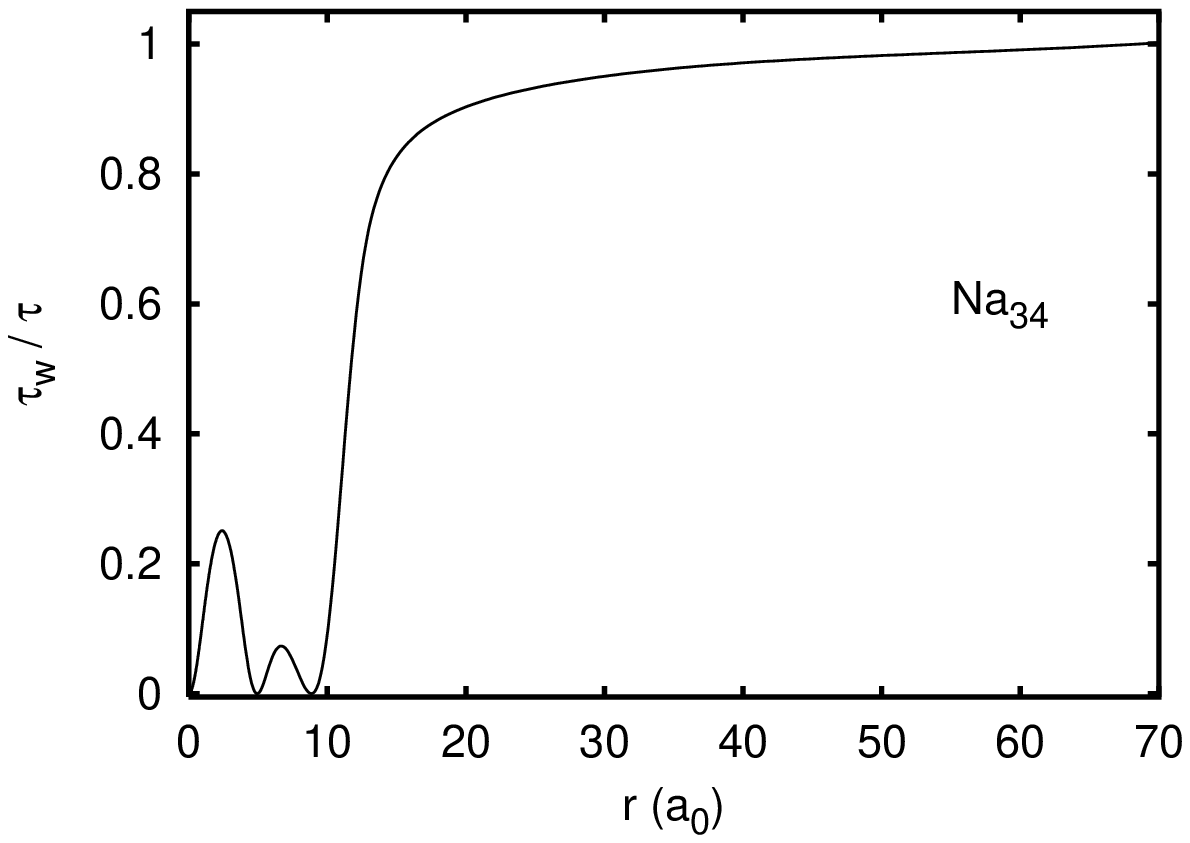}
\caption{The kinetic-energy-based iso-orbital indicator
$x(\re)=\tau^W(\re)/\tau(\re)$ of Eq.\ (\ref{defx}), as a function of
the radial coordinate $r$ (in bohr) for two spherical densities: the
neon atom, and a jellium cluster resembling $\mathrm{Na}_{34}$, with a
radius of 12.7 bohr.
\label{fig1}} 
\end{figure}
Fig.\ \ref{fig1} shows $x_\sigma(\re)$  of Eq.\ (\ref{defx}), plotted
as a function of the distance from the center of two spherical
densities, the Ne atom and a jellium cluster with 34 electrons having
the average valence electron density of Na 
($r_s=[3/(4\pi n)]^{1/3}=3.93$ bohr). $x_\sigma(\re)$ correctly
identifies the density tails as iso-orbital. In the Ne atom, the 1s
core is also found to be nearly iso-orbital. The interior of the
jellium cluster, however, is found to be a region of strong orbital
overlap, as expected. The densities and orbitals have been evaluated
by solving \cite{talman,oep} the OEP problem for exact exchange.

While $x_\sigma(\re)$ of Eq.\ (\ref{defx}) seems to be a satisfactory
iso-orbital indicator, it does display an order-of-limits problem
\cite{newmgga}: Define 
$\alpha=(\tau-\tau^W)/\tau_0$, where 
$\tau_0=\frac{3 \hbar^2}{10m}(3\pi^2)^{2/3} n^{5/3}$, and
$p=\frac{3}{5}(\tau^W/\tau)=|\nabla n|^2/[4 (3 \pi^2)^{2/3}
n^{5/3}]$. Then, for $\nup=\ndo$,
\be
x=\frac{\tau^W}{\tau}=\frac{1}{1+\frac{3}{5}\frac{\alpha}{p}},
\ee
\be
\lim_{p\rightarrow 0}\lim_{\alpha\rightarrow 0} x=1
\hspace{4mm}\mathrm{but} \hspace{4mm}
\lim_{\alpha\rightarrow 0}\lim_{p\rightarrow 0} x=0.
\ee
This problem shows up in nearly-iso-orbital ($\alpha \rightarrow 0$)
regions where the gradient of the density approaches zero ($p
\rightarrow 0$), and thus perhaps at covalent bond centers.

\begin{figure}[t]
\includegraphics[width=8.1cm]{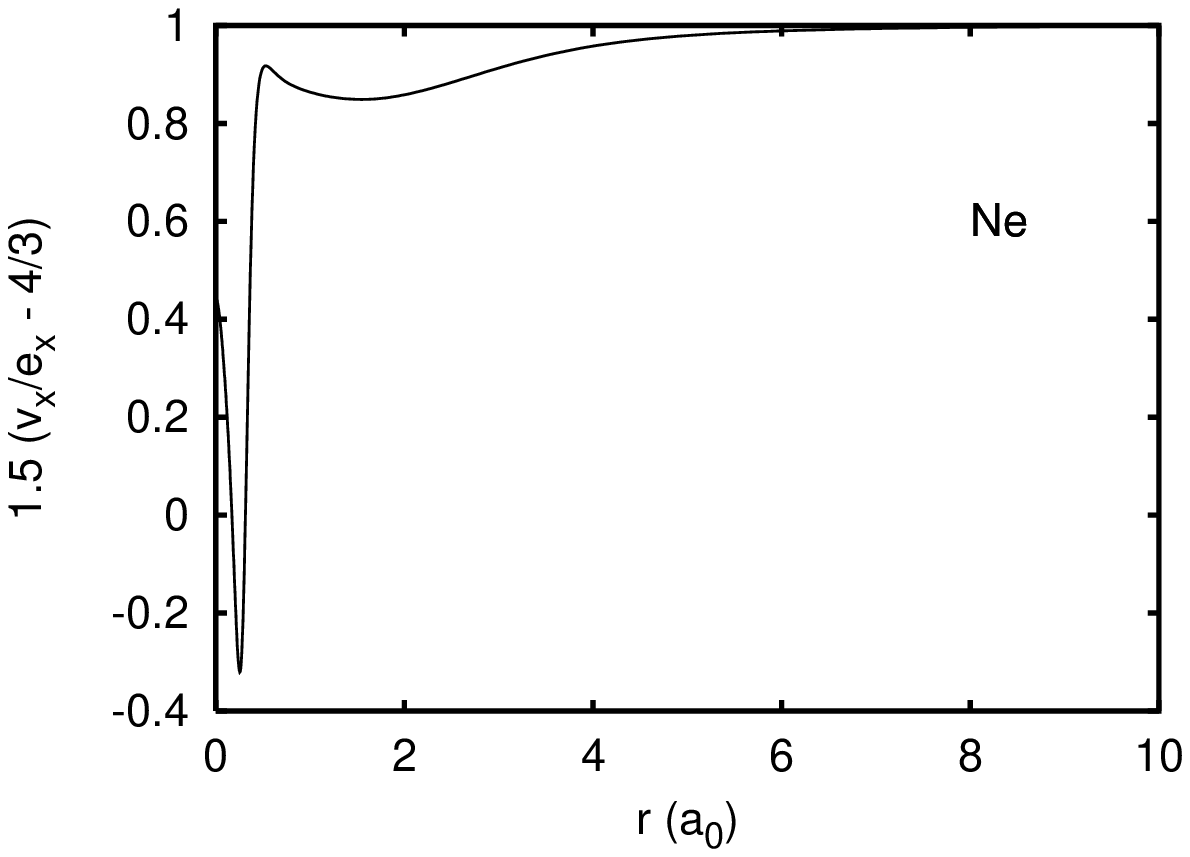}
\includegraphics[width=8.1cm]{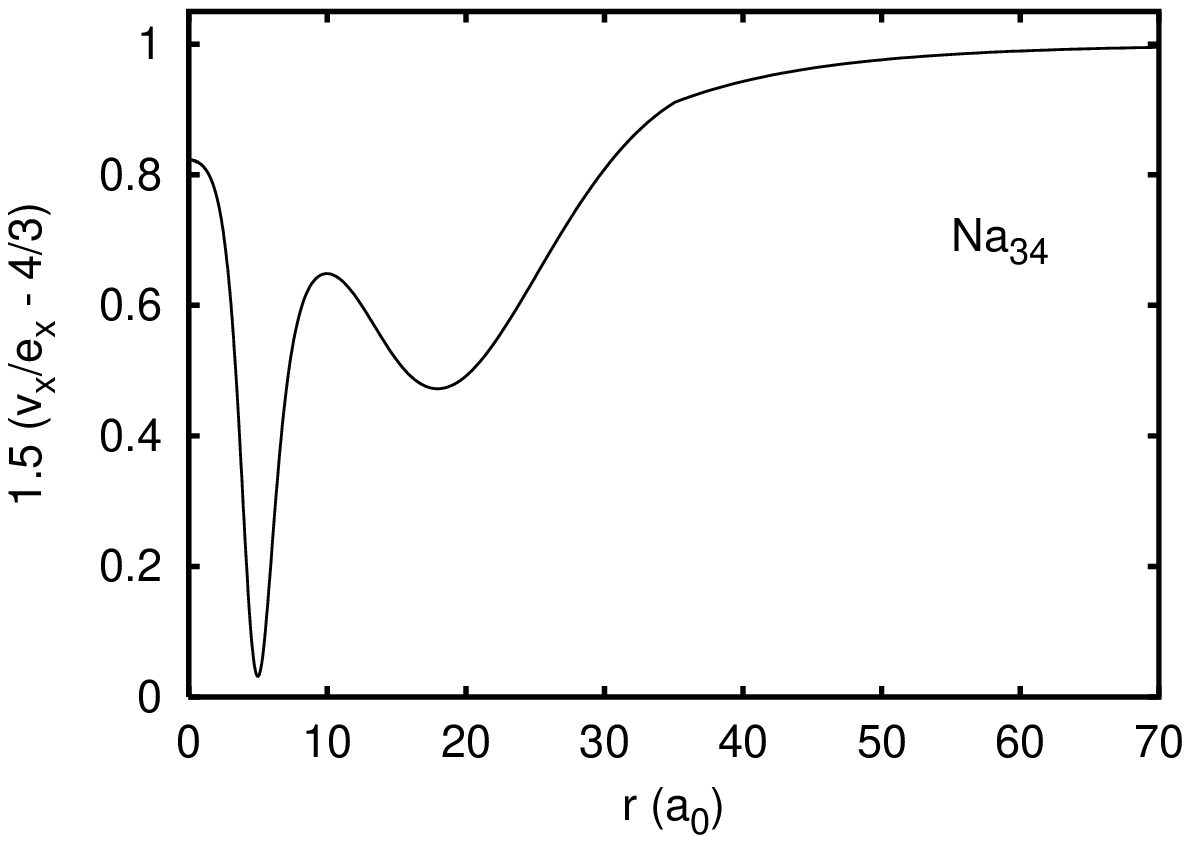}
\caption{The exchange-based iso-orbital indicator $\tilde{x}=\frac{3}{2}\left(
\frac{v_\mathrm{x}(\re) }{e_\mathrm{x}(\re)}-\frac{4}{3}\right)$ of
Eq.\ (\ref{tilx}), for the same densities as in Fig.\ (\ref{fig1}).
\label{fig2}} 
\end{figure}

\begin{table}[b]
\begin{center}
\begin{tabular}{|c|c||c|c|}
\hline
N & ratio &
N & ratio \\ \hline
2   & 1.000 & 40  & 0.547 \\ \hline
8   & 0.834 & 58  & 0.528 \\ \hline
18  & 0.679 & 92  & 0.528 \\ \hline
20  & 0.640 & 106 & 0.390 \\ \hline
34  & 0.571 & 138 & 0.442 \\ \hline
\end{tabular}
\caption{The ratio of Eq.\ (\ref{ixtil}) for
closed-shell jellium spheres of increasing electron number N.
\label{tab1}
} 
\end{center}
\end{table}
Because of the order-of-limits problem of $\tau^W/\tau$, it may be
worthwhile to consider alternative iso-orbital indicators. For example,
the exact exchange potential $v_\mathrm{x}(\re)$ and the exact
exchange energy per electron $e_\mathrm{x}(\re)$ (Eq.\ (\ref{ex})) of
a spin-unpolarized system are related in the iso-orbital limit by
\be
\frac{v_\mathrm{x}(\re)}{e_\mathrm{x}(\re)}=2 \hspace{5mm} (N=2),
\ee
and in the uniform-density limit by 
\be
\frac{v_\mathrm{x}(\re)}{e_\mathrm{x}(\re)}=\frac{4}{3} \hspace{5mm}
(\mbox{uniform density}).
\ee
One might define 
\be
\label{tilx}
\tilde{x}(\re)=\frac{3}{2}\left(
\frac{v_\mathrm{x}(\re) }{e_\mathrm{x}(\re)}-\frac{4}{3}\right)
\ee
as an alternative iso-orbital indicator, which varies from 1 in the
iso-orbital limit to 0 in the uniform limit. Table \ref{tab1} shows
that
\be
\label{ixtil}
\frac{3}{2}\left[
\frac{\int d^3r \, n(\re) v_\mathrm{x}(\re) }
{\int d^3r \, n(\re) e_\mathrm{x}(\re)}-\frac{4}{3}\right]
\ee
varies almost smoothly from 1 for the $N=2$ jellium cluster to 0.4 for
the largest cluster studied here. Fig.\ \ref{fig2} however shows that
$\tilde{x}(\re)$ of Eq.\ (\ref{tilx}) can be negative, fails to
recognize the 1s core of the Ne atom as a strongly iso-orbital region,
and fails to recognize the interior of the jellium cluster as a region
of strongly overlapped orbitals.

Thus it seems that $x_\sigma(\re)$ of Eq.\ (\ref{defx}) is the
preferred iso-orbital indicator. Note that Eq.\ (\ref{tauw}) assumes
that the orbitals can be chosen to be real. If the state of interest
has a nonzero current density 
\be
\mathbf{j}_\sigma(\re)= \mathrm{Re}\, \frac{\hbar}{m}
\sum_\alpha^\mathrm{occup.} \varphi_{\alpha\sigma}^*(\re)
\frac{1}{i} \nabla \varphi_{\alpha\sigma}(\re),
\ee
then \cite{dobson,becke2}
\be
\tau^W_\sigma(\re)=\frac{\hbar^2}{8m}
\frac{|\nabla n_\sigma(\re)|^2}{n_\sigma(\re)}
+\frac{m|\mathbf{j}_\sigma(\re)|^2}{2n_\sigma(\re)}.
\ee
To ensure that $x_\sigma$ equals one for a one-electron density and
zero for a uniform density (with or without a uniform current), the
best choice may be
$x_\sigma=\tilde{\tau}_\sigma^W/\tilde{\tau}_\sigma$ where
$\tilde{\tau}_\sigma^W=\tau^W_\sigma-m|\mathbf{j}_\sigma|^2/(2n_\sigma)$
and  
$\tilde{\tau}_\sigma=\tau_\sigma-m|\mathbf{j}_\sigma|^2/(2n_\sigma)$.
In this way, the self-correlation error can be corrected even in a general
excited state.

\section{Conclusions}

For many standard applications of ground-state density functional
theory, the self-interaction errors of modern GGA's and meta-GGA's are
relatively benign. There are a few striking exceptions to this rule,
such as the binding properties of diatomic molecules with an odd
number of valence electrons \cite{jp,gruen} and the static
(hyper-) polarizabilities of long-chain molecules
\cite{gisbergen}. For applications involving time-dependent and
excited-state Kohn-Sham density functional theory
\cite{chu2,petersilka}, the self-interaction errors can be severe.

While the Perdew-Zunger self-interaction correction to the local spin
density approximation can now be brought under the umbrella of
Kohn-Sham theory, the development of more sophisticated functionals
and optimized effective potential methods suggests that
general-purpose self-interaction-free density functionals will be
developed soon and implemented within Kohn-Sham theory. Such
functionals may well include full exact exchange plus highly nonlocal
correlation based in part upon unitarily-invariant iso-orbital
indicators such as $\tau^W_\sigma(\re)/\tau_\sigma(\re)$.

\vspace{3mm}
Acknowledgements: Our investigation of Eq.\ (\ref{tilx}) was triggered
by discussions with Prof.\ Dietmar Kolb. S.K.\ acknowledges financial
support by the Deutsche Forschungsgemeinschaft under an 
Emmy-Noether grant, and J.P.P.\ by the U.S.\ National Science Foundation 
under grant DMR 01-35678.

\end{document}